\documentclass[final,3p,times,twocolumn]{elsarticle} 

\usepackage{amssymb}
\usepackage{adjustbox}
\usepackage{lmodern}
\usepackage{url}
\usepackage{hyperref}

\usepackage{lineno}

\journal{Nuclear Instruments and Methods in Physics Research A}

\begin{document}

\begin{frontmatter}



\title{Experimental techniques to study the $\gamma$ process for nuclear astrophysics at the Cologne accelerator laboratory}


\author[col]{F.~Heim\corref{cor1}}
\ead{fheim@ikp.uni-koeln.de}
\cortext[cor1]{Corresponding author}
\author[col]{J.~Mayer}
\author[col]{M.~M\"uller}
\author[nd]{P.~Scholz}
\author[col]{M.~Weinert}
\author[col]{A.~Zilges}

\address[col]{University of Cologne, Institute for Nuclear Physics, Z\"ulpicher Stra{\ss}e 77, D-50937 Cologne, Germany}
\address[nd]{University of Notre Dame, Department of Physics, Indiana 46556-5670, USA}

\begin{abstract}
The nuclear astrophysics setup at the Institute for Nuclear Physics, University of Cologne, Germany is dedicated to measurements of total and partial cross sections of charged-particle induced reactions at astrophysically relevant energies. These observables are key ingredients for reaction network calculations of various stellar scenarios, and crucial for the understanding of the nucleosynthesis of elements. The experiments utilize the high-efficiency $\gamma$-ray spectrometer HORUS, and the 10 MV FN-Tandem accelerator.
An updated target chamber as well as further experimental methods established in the last years will be presented which allow to measure cross sections down to the nb region. The reliability of the measured cross sections is proven by a $^{89}$Y(p,$\gamma$)$^{90}$Zr commissioning experiment.
Additionally, an application for nuclear astrophysics will be presented. The results of a $^{93}$Nb(p,$\gamma$)$^{94}$Mo experiment will be discussed as well as their deviations compared to formerly reported results.

\hspace*{0.2mm}

\noindent Published in Nuclear Instruments and Methods in Physics Research Section A 966, 163854 (2020). 

\noindent \href{https://doi.org/10.1016/j.nima.2020.163854}{DOI: 10.1016/j.nima.2020.163854} \copyright 2020. This manuscript version is made available under the CC-BY-NC-ND 4.0 license \url{http://creativecommons.org/licenses/by-nc-nd/4.0/}.
\end{abstract}




\end{frontmatter}


\section{Introduction}
\label{sec:introduction}
The longstanding question why, and in particular how different elements are formed inspired the creation of the interdisciplinary field of nuclear astrophysics \cite{bbfh}. The correct description of the complex processes found in various astrophysical scenarios typically requires a detailed understanding of the underlying
nuclear physics. In particular, the nucleosynthesis of neutron deficient $p$ nuclei remains one of the unsolved puzzles \cite{Arnould2003, Rauscher2013}. The $\gamma$ process, which is assumed to be responsible for the largest contribution to the abundance of the $p$ nuclei, builds a huge network of photodisintegration reactions and includes thousands of
different reactions on mainly unstable and exotic nuclei. Thus, one often needs to rely on theoretical calculations for estimates of cross-sections of reactions away from the experimentally known regions. The extension of the available experimental database is therefore one of the main tasks of experimental nuclear astrophysics. Since cross sections at astrophysically relevant energies, \textit{i.e.}, inside the Gamow window, are typically in the lower $\mu$b range, sensitive measurement techniques are mandatory.

Various direct methods are available for the determination of absolute cross sections.
A well-established technique for cross-section measurements is the \emph{activation method}. This is a two-step method, during the first step unstable reaction products are produced and during the second step the radioactive decay is analyzed in a counting setup. In most cases the $\gamma$-ray transitions in the daughter nucleus are observed but $\alpha$- or $\beta$-particles might also be counted. This method requires unstable reaction products with appropriate decay half-lives and decay schemes. A review article on the activation method can be found in Ref. \cite{Gyurky2019} and recent experimental data are provided, e.g., in Refs. \cite{Netterdon2014, Gyurky2014, Gyuky2014_2, Kiss2014, Scholz2014, Guray2015, Kiss2015, Yalcin2015, Halasz2016, Kinoshita2016, Scholz2016, Ozkan2017}.

The \emph{in-beam 4$\pi$-summing technique} overcomes some limitations of the activation technique and utilizes a large scintillator crystal which covers a solid angle of almost 4$\pi$ around the target position and summarizes the energies of all $\gamma$-rays emitted in a certain time window \cite{Spyrou07, Spyrou08, Simon13, Simon19, reingold19}. 

\begin{figure}[t]
\centering
\includegraphics[width=0.9\columnwidth]{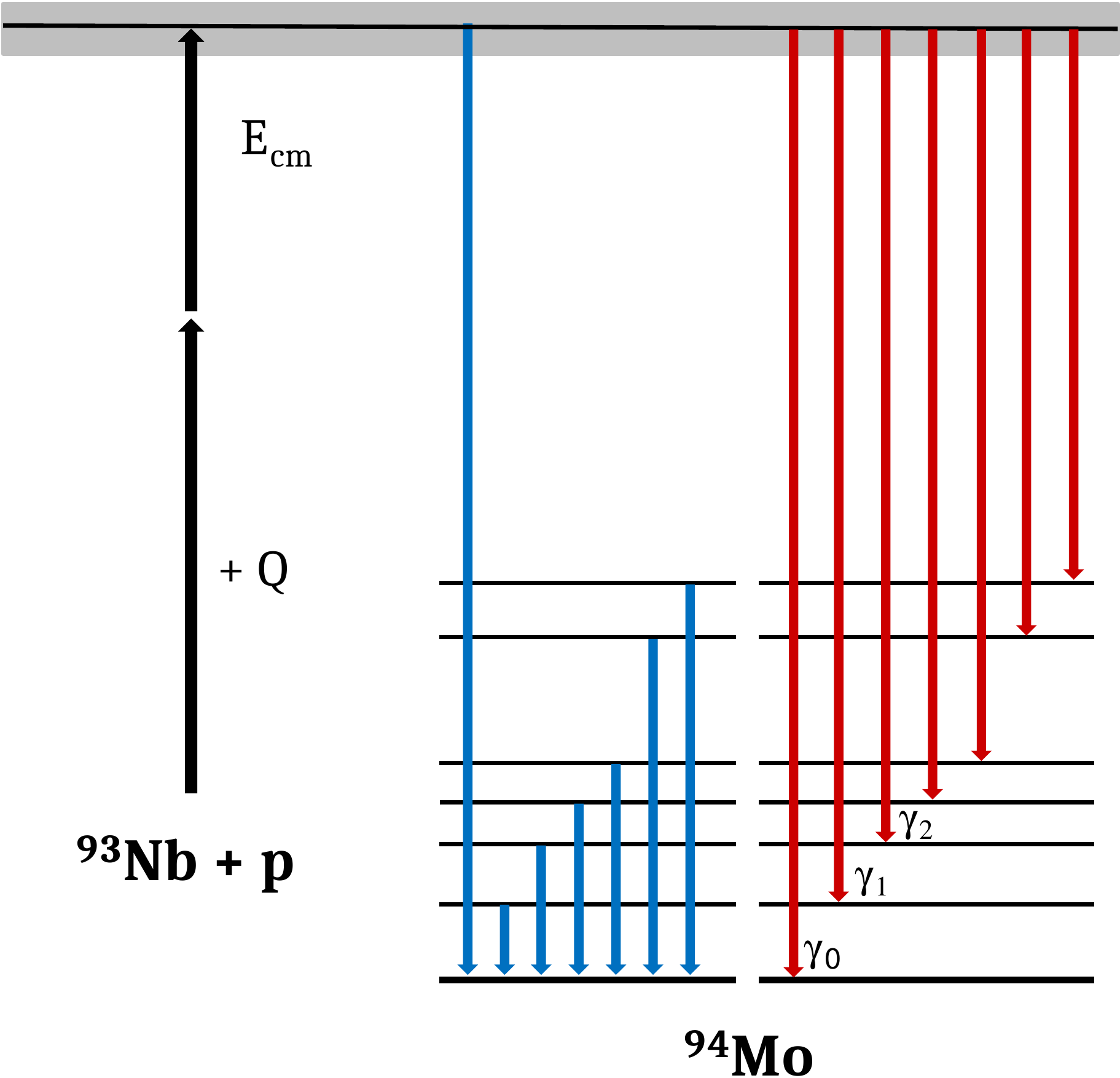}
\caption[Illustration of the in-beam technique and reaction mechanism]{Illustration of the reaction mechanism that leads to the highly excited compound nucleus after a radiative-capture reaction. The intensity for the prompt de-excitation into the ground state of the reaction product directly ($\gamma_0$) or into various levels ($\gamma_1$, $\gamma_2$ and so on) is called a \emph{partial cross section} (red arrows). The sum of all ground-state transitions is used to determine the \emph{total cross section} (blue arrows).}  
\label{fig:method}
\end{figure}

Measuring radiative-capture reaction cross sections with both stable and unstable reaction products is also feasible with the \emph{in-beam high resolution $\gamma$-ray spectroscopy technique}. The basic idea of this method is the observation of the prompt decay via $\gamma$-ray transitions from a highly excited compound nucleus with an excitation energy of $E_x=Q+E_{cm}$ into different states of the reaction product (see Fig. \ref{fig:method}), where $E_{cm}$ denotes the center-of-mass energy. Note that other than radiation emitted from a source, the photons stemming from the decay of the excited compound nucleus are not emitted isotropically but with an angular distribution with respect to the beam axis. Utilizing a multi detector $\gamma$-ray spectrometer in combination with a dedicated target chamber allows to measure these angular distributions, and hence to extract absolute cross sections. 

This method and the dedicated setup for the in-beam measurement of absolute cross sections at the Tandem accelerator lab of the University of Cologne is addressed in this paper.
In Section \ref{sec:horus} experimental details of the in-beam $\gamma$-ray spectroscopy method will be presented. In particular, emphasis will be given to recent developments which secure a more efficient and reliable analysis of cross-section experiments. In Section \ref{sec:chamber} details of the revised target chamber dedicated for the experiments will be presented as well as results of the  $^{89}$Y(p,$\gamma$)$^{90}$Zr commissioning experiment. The results of the astrophysical relevant experiment $^{93}$Nb(p,$\gamma$)$^{94}$Mo will be discussed in Section \ref{sec:experiments}.

\section{$\gamma$-ray spectroscopy at HORUS in Cologne}
\label{sec:horus}

The $\gamma$-ray spectrometer HORUS (\textbf{H}igh efficiency \textbf{O}bservatory for $\gamma$-\textbf{R}ay \textbf{U}nique \textbf{S}pectroscopy) is a multi-purpose setup which is used for $\gamma$-ray spectroscopy experiments addressing very different purposes. Various target chambers dedicated to special types of experiments can be installed inside HORUS, see e.g.,  Refs. \cite{Scholz2016, Netterdon_Sn, Hennig15, pickstone2017, Spieker18}.

\begin{figure}[t]
\centering
\includegraphics[width=1\columnwidth]{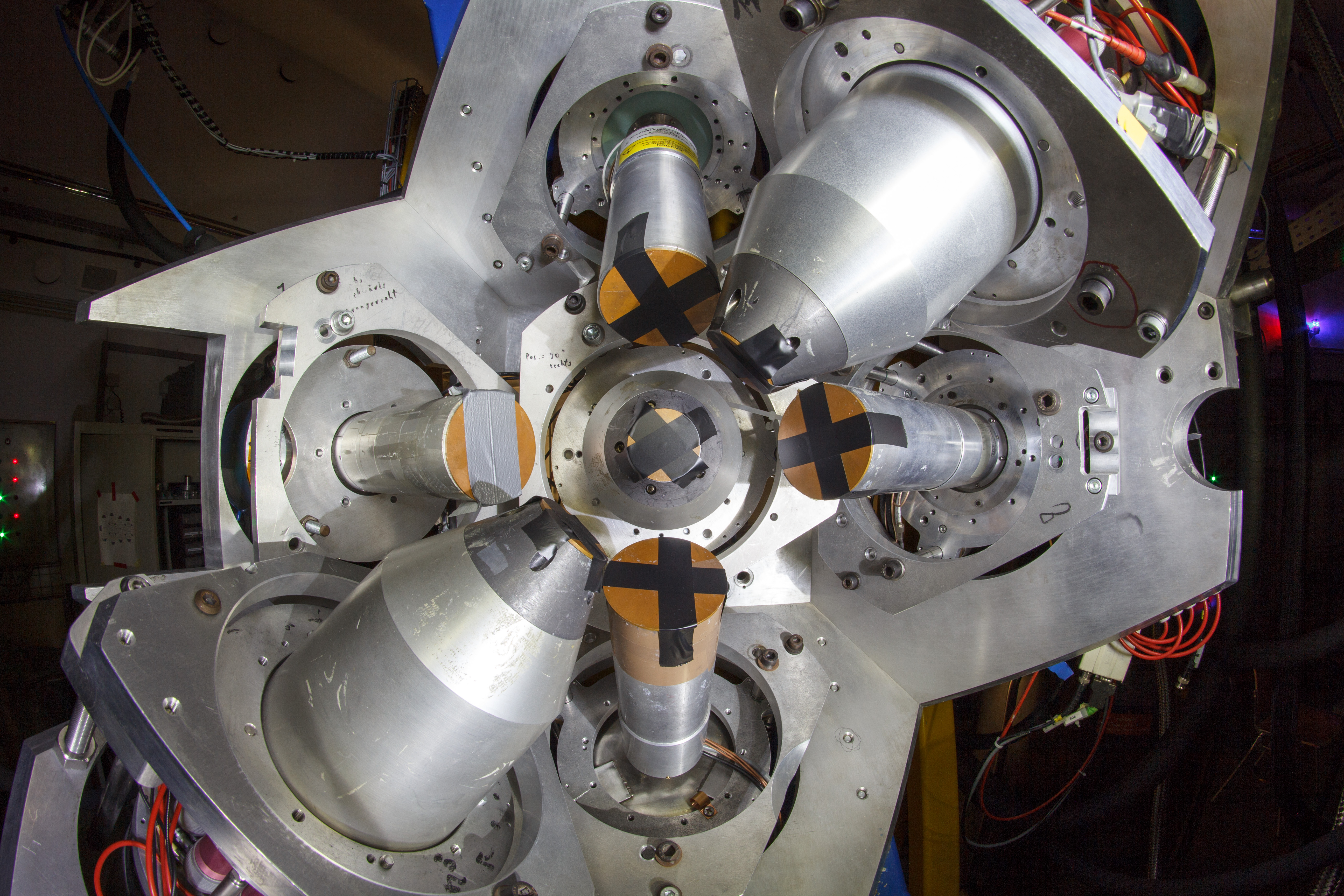}
\caption[Illustration of the in-beam technique and reaction mechanism]{One hemisphere of the HORUS $\gamma$-ray spectrometer. The target chamber is installed in the center at the point of intersections of all 14 HPGe detectors. Six detectors can be equipped with BGO shields for an active Compton background suppression.}  
\label{fig:horus}
\end{figure}

HORUS can hold up to 14 High Purity Germanium (HPGe) detectors of which six can be equipped with Bismuth germanate (BGO) shields for an active suppression of background from Compton scattering. Its geometry is based on a cube with 14 HPGe detectors placed on its six faces and eight corners. This leads to a coverage of five different angles of 35$^{\circ}$, 45$^{\circ}$, 90$^{\circ}$, 135$^{\circ}$ and 145$^{\circ}$ with respect to the beam axis, which allows the determination of angular distributions. One hemisphere of the HORUS spectrometer is shown in Fig. \ref{fig:horus}. All detector end caps can be shielded with copper and lead plates of thicknesses of up to a few millimeters to suppress low energy $\gamma$- and X-rays and the detectors are positioned as close to the target chamber as possible, typically at distances of 9 cm up to 17 cm.

The preamplified signals from the HPGe detectors are processed using the Digital Gamma Finder (DGF-4C Rev. F) modules by XIA \cite{Hubbard-Nelson99, Skulski00}. Each module provides four input channels as well as dedicated VETO inputs for each channel, which are used for
the active background suppression of the BGO shields. The incoming signals are digitized at a rate of 80 MHz by flash ADCs with a depth of 14 bit. Detector, energy and timing information are stored in a listmode format, allowing an offline analysis of the data including $\gamma\gamma$ coincidences \cite{Hennig14}. The dynamic energy range is adjustable. For cross-section studies the upper energy limit is usually set to 12-18 MeV.

\subsection{Beam-energy calibration}
\label{subsec:calib}

The $\gamma$-process nucleosynthesis is assumed to appear in explosive stellar scenarios at temperatures of about 2.0 to 3.0 GK \cite{Rauscher2013}. The Maxwell-Boltzmann distribution provides a velocity distribution for the interacting particles in the plasma. Since the tunneling probability and hence the cross section increases exponentially with increasing particle energy, the astrophysically relevant energy region -- the Gamow window -- is given by a convolution of those two probability functions, see e.g. Ref. \cite{Rauscher2010}. For this reason, a precise determination of the beam energy in astrophysically motivated experiments is inevitable. 

While passing through the target material the beam particles will loose energy $\Delta E$ before leaving the target layer and reaching the gold backing foil. This loss is estimated using the SRIM simulation code \cite{Srim}. The effective beam energy is determined by:
\begin{equation} \label{eq:ep}
E_p=E_{NMR}+E_{OS}-\frac{\Delta E}{2},
\end{equation}
where $E_p$ denotes the effective proton beam energy, $E_{NMR}$ the beam energy expected from the settings of the Nuclear Magnetic Resonance (NMR) probe inside the 90$^{\circ}$ deflecting magnet of the Tandem accelerator and $E_{OS}$ the observed offset of the NMR probe. 
In the following two different approaches to determine the energy offset will be presented, as well as their specific advantages and disadvantages.

\begin{figure}[t]
\centering
\includegraphics[width=1\columnwidth]{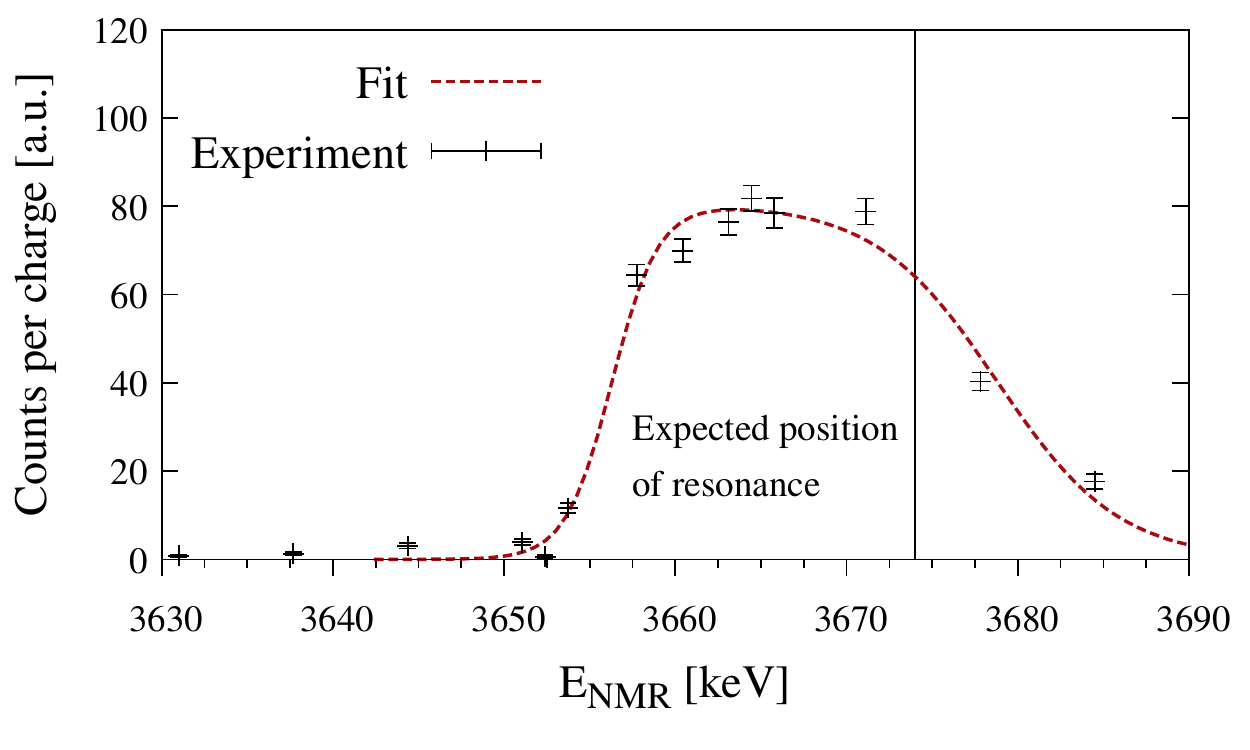}
\caption{Resonance curve of the $^{27}$Al(p,$\gamma$)$^{28}$Si reaction on a 100 $\mu$g/cm$^2$ thick Al target. The results show that the actual beam energy is about 17 keV higher than expected from the settings of the NMR probe inside the 90$^{\circ}$ deflecting magnet of the Tandem accelerator. This resonance scan is an appropriate method to determine the beam energy, but is time-consuming since it requires at minimum one day of beam-time.}  
\label{fig:resonance}
\end{figure}

\subsubsection{The $^{27}$Al(p,$\gamma$)$^{28}$Si resonance}
\label{subsub:27Al}
The resonance of the $^{27}$Al($p,\gamma$)$^{28}$Si reaction at $E_p$ = 3674.4 keV \cite{Brenneisen1995} was scanned in small energy steps (up to a few keV). The peak volume of the 2838 keV peak, \emph{i.e.}, the transition $4_1^+ \rightarrow 2_1^+$ in $^{28}$Si, was normalized to the accumulated charge deposited on the target. The result is a resonance curve, which is shown in Fig \ref{fig:resonance}. The width of the rising edge of the resonance curve is determined by the energy spread of the proton beam provided by the Tandem accelerator and amounts to about 8 keV. The width of the plateau is caused by the energy loss of the protons in the Al target. From the position of the rising edge the offset, and hence the effective particle beam energy can be determined.


\begin{figure}[t]
\centering
\includegraphics[width=\columnwidth]{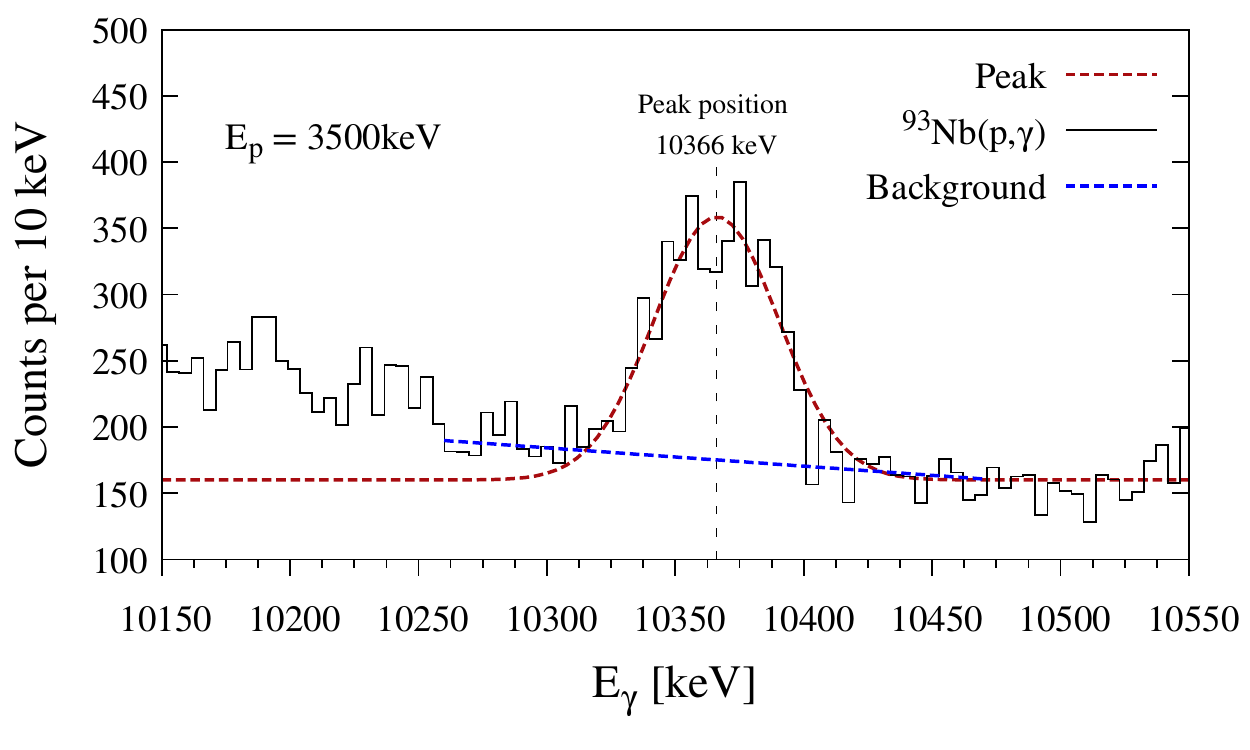}
\caption{The shown peak contains primary $\gamma$-ray transitions that directly populate the 2$^{nd}$ excited state in $^{94}$Mo at 1574 keV. The $Q$ value of the $^{93}$Nb($p,\gamma$)$^{94}$Mo reaction is $Q$=8490 keV. The energy loss $\Delta$E of the protons amounts to about 49 keV and causes a broad peak, however this method allows an energy calibration of the accelerator with in-beam $\gamma$-ray spectra taken from the experiment.}  
\label{fig:prompt}
\end{figure}

\subsubsection{Beam calibration from prompt $\gamma$-rays}
\label{subsub:prompt}

In most of the radiative-capture experiments we are interested in the prompt $\gamma$-ray de-excitation of the compound state as indicated by red arrows in Fig. \ref{fig:method}. From the position of the respective primary $\gamma$-ray transitions the center-of-mass energy can be deduced via:
\begin{equation} \label{eq:gamma}
E_{c.m.}=E_\gamma-Q+E_{state},
\end{equation}
where $E_\gamma$ is the $\gamma$-ray energy of the prompt transition, $E_{state}$ is the respective energy of the state that is populated by the primary $\gamma$-ray and $Q$ is the $Q$ value of the reaction. For detectors with an angle of 90 degrees with respect to the beam axis the combination of Eq. \ref{eq:ep} and \ref{eq:gamma} directly yields the energy offset of the accelerator from the in-beam $\gamma$-ray spectra. For angles different from 90$^{\circ}$, Doppler corrections need to be taken into account. As shown in Fig. \ref{fig:prompt} the corresponding peaks are broadened due to the energy loss in the target, but with an appropriate amount of statistics the peak positions can be extracted very accurately.
The main advantage of this method is that no modification of the experimental setup, \textit{i.e.}, variation of beam energy and/or target, is required. The uncertainty of the energy determination using this method is mainly defined by the uncertainty of the energy loss in the target as well as by the precision of the peak position. The width of the prompt peak in Fig. \ref{fig:prompt} is about 23 keV. The uncertainty of the energy loss is almost negligible, since the target thickness can be determined quite accurately (see Sec. \ref{subsec:target}). Finally, as an total uncertainty for the determination of the beam energy we approximate 25 keV for the data shown in Fig. \ref{fig:prompt}.

\subsection{Angular correlations}
\label{subsec:angular}

\begin{figure}[t]
\centering
\includegraphics[width=1\columnwidth]{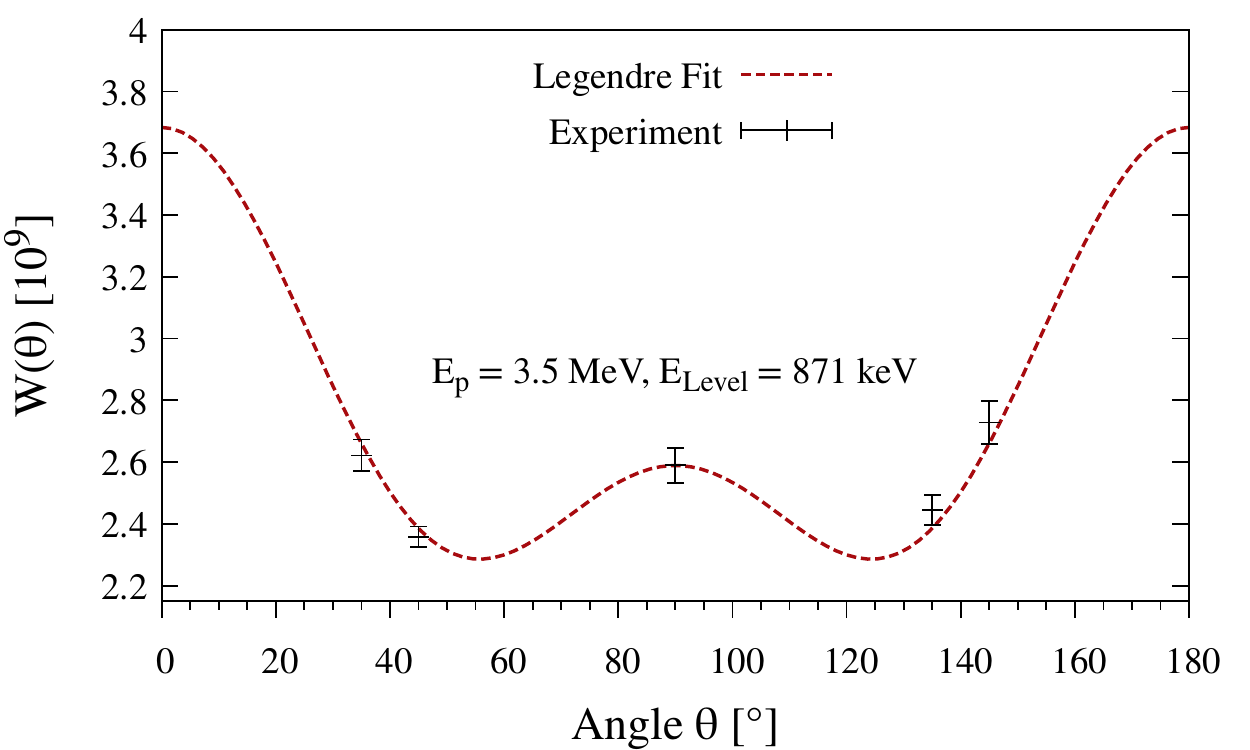}
\caption{The radiation emitted from the excited compound nucleus follows an angular distribution. The experimentally determined values for the individual angles can be fitted by a series of Legendre polynomials. Here shown as an example is the $2_1^+ \rightarrow g.s.$ in $^{94}$Mo ($E_{\gamma}$~=~871 keV) for an incident proton energy of 3.5 MeV.}  
\label{fig:angular}
\end{figure}

Although compound reactions are the dominant reaction mechanism for particle energies of astrophysical interest, a contribution of memory-preserving, direct-like reaction processes is exhibited by the measured angular distributions. Note that the measured angular distribution is a superposition of contributions from several $\gamma$-ray cascades. The experimental yield $Y(E_{\gamma})$ is first corrected for the full-energy peak efficiency $\epsilon(E_{\gamma})$ and the dead time correction of the data acquisition $\tau$:
\begin{equation} 
W(\Theta)=\frac{Y(E_{\gamma},\Theta)}{\epsilon(E_{\gamma})\cdot \tau}
\end{equation}
At the HORUS $\gamma$-ray spectrometer $Y(E_{\gamma})$ is measured at five angles and the angular distribution is obtained by fitting a sum of Legendre polynomials to the five experimental values:
\begin{equation} \label{eq:angular}
W(\Theta)=A_0\left(1+\sum_{k=2,4} \alpha_kP_k(cos\Theta)\right)
\end{equation}
In this equation A$_0$ and $\alpha_k$ denote energy dependent coefficients and P$_k$ the Legendre polynomials P$_2$ and P$_4$. Taking only Legendre polynomials up to the order of $k=~4$ into account is justified by the assumption that dipole and quadrupole transitions dominate the electromagnetic de-excitations. Recent experiments have shown that $\alpha_k$ varies slightly with beam energy \cite{Mayer2016, Heim2019, Scholz2019}. Hence, the angular distributions need to be obtained for each $\gamma$-ray transition at each beam energy. Figure \ref{fig:angular} shows an example of the angular distribution for the decay $2_1^+ \rightarrow g.s.$ in $^{94}$Mo ($E_{\gamma}$ = 871 keV) for an incident proton energy of 3.5 MeV.

\subsection{Full-energy peak efficiency}
\label{subsec:effi}

\begin{figure}[t]
\centering
\includegraphics[width=1\columnwidth]{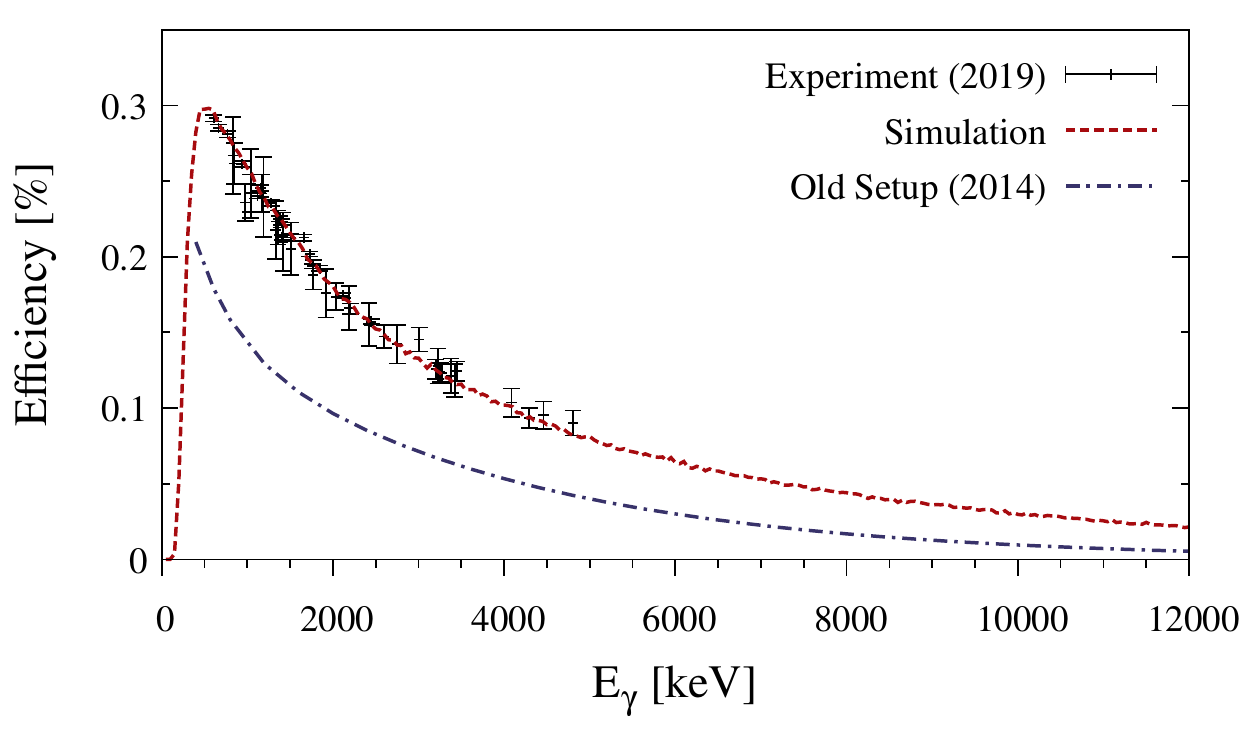}
\caption{The shown full-energy peak efficiencies were extracted using a standard calibration $^{226}$Ra source as well as in-house produced $^{56}$Co (T$_{1/2}$=77\,d) and $^{66}$Ga (T$_{1/2}$=9.5\,h) sources for one HPGe detector at a distance of 11 cm. The in-house produced sources allow to determine experimental efficiencies up to $\gamma$-ray energies of 4.8 MeV. The simulated efficiencies using \textsc{Geant4} \cite{g4Horus} agree very well with the experimental results. For comparison, the efficiency using the old target chamber are shown in dash-dotted (see. Sec. \ref{sec:chamber} for details). Efficiencies for the same detector at the same distance are shown.}  
\label{fig:effi}
\end{figure}

Full-energy peak efficiencies are required for the determination of absolute cross sections and are typically obtained using the standard calibration sources $^{152}$Eu and $^{226}$Ra. They provide absolute efficiencies up to $\gamma$-ray energies of about 2.5 MeV. An in-house produced $^{56}$Co source (T$_{1/2}$=77\,d) emits photons of energies up to 3.5 MeV. The relative efficiencies can be normalized to data from the calibrated sources. However, for many experiments the energy range needs to be extended further up to about 10 MeV. 

In the past, the $^{27}$Al(p,$\gamma$)$^{28}$Si resonance at $E_p$ = 3674.4 keV (see \mbox{Sec. \ref{subsec:calib} \ref{subsub:27Al})} has been used frequently. The absolute $\gamma$-decay branching ratios for the depopulaton of the excited state at $E_x$ = 15\,127 keV are known from Ref. \cite{Brenneisen1995}. However, all $\gamma$-rays stemming from this resonance follow an angular distribution. The coefficients $\alpha_2$ and $\alpha_4$ are also determined in Ref.  \cite{Brenneisen1995} but have large uncertainties. 
This large error makes the results less useful. The transitions into higher-lying states with $\gamma$-ray energies of about 4-6 MeV, which are used to scale the efficiencies, exhibit lower uncertainties for $\alpha_2$ and $\alpha_4$ but their statistical uncertainties amount to up to 25 \%. Therefore, the absolute efficiencies determined from the $^{27}$Al(p,$\gamma$)$^{28}$Si resonance may be doubtful.

A new approach to extend the efficiency calibrated energy range is the use of an in-house produced $^{66}$Ga source. The rather short half-life of 9.5\,h demands to perform the $^{66}$Zn(p,n) activation immediately before the calibration experiment. This activation demands for proton beam energies of about 10 MeV, beam intensities of a few nA and irradiation times of only 1-2 hours. $^{66}$Ga emits several high-intensity $\gamma$-rays of energies between 3.5 and 4.8 MeV \cite{66Ga} and it can be normalized accurately to data from, \emph{e.g.}, a $^{226}$Ra source at lower energies. Note, that the two highest $\gamma$-ray transitions at 4806 keV and 4295 keV are exactly 511 keV apart, and hence single-escape contributions are buried in the full-energy peak at 4295 keV. There are two different ways to compensate for this:

First, if the experimental setup is sufficiently well simulated, \emph{e.g}, with \textsc{Geant4} \cite{Geant4}, not only full-energy peak efficiencies (see Fig. \ref{fig:effi}) but also single-escape efficiencies can be obtained. Hence, the single-escape contribution can be estimated.

Secondly, the $\gamma$-ray at 4295 keV describes a ground-state transition in $^{66}$Zn. The corresponding level also de-excites via other $\gamma$-ray transitions with very precisely known branching ratios. In particular, the 1190~keV transition into the $0_4^+$ state is well-suited to estimate the intensity of the 4295 keV $\gamma$-ray transition. In our experiments both of these approaches delivered very similar results with relative uncertainties of about 10 \%. The full-energy peak efficiencies obtained from $^{226}$Ra, $^{56}$Co and $^{66}$Ga sources are fitted by a function of the form
\begin{equation} \label{eq:fit}
\epsilon(E_\gamma)=a\cdot exp(-b\cdot E_\gamma)+c\cdot exp(-d\cdot E_\gamma).
\end{equation}
The fit parameters are given in Tab. \ref{tab:fit}.

\begin{table}[t]
\centering
\caption{Fit parameter for the efficiency function given in Eq. \ref{eq:fit}. The total detection efficiency at $E_\gamma$ = 1300 keV amounts to about 3.3 \%.}
\vspace{7pt}
\begin{adjustbox}{width=1\columnwidth}
\begin{tabular}{cccc}
a & b [keV$^{-1}$] &c & d [keV$^{-1}$]   \\ \hline \hline
4.8(3)$\times 10^{-2}$ & 3.2(2)$\times 10^{-4}$	&5.5(4)$\times 10^{-2}$ & 3.0(2)$\times 10^{-3}$  \\ \hline
\end{tabular}
\end{adjustbox}
\label{tab:fit}
\end{table}

\subsection{$\gamma$-$\gamma$ coincidence measurements}
\label{subsec:gammagamma}

\begin{figure}[t]
\centering
\includegraphics[width=1\columnwidth]{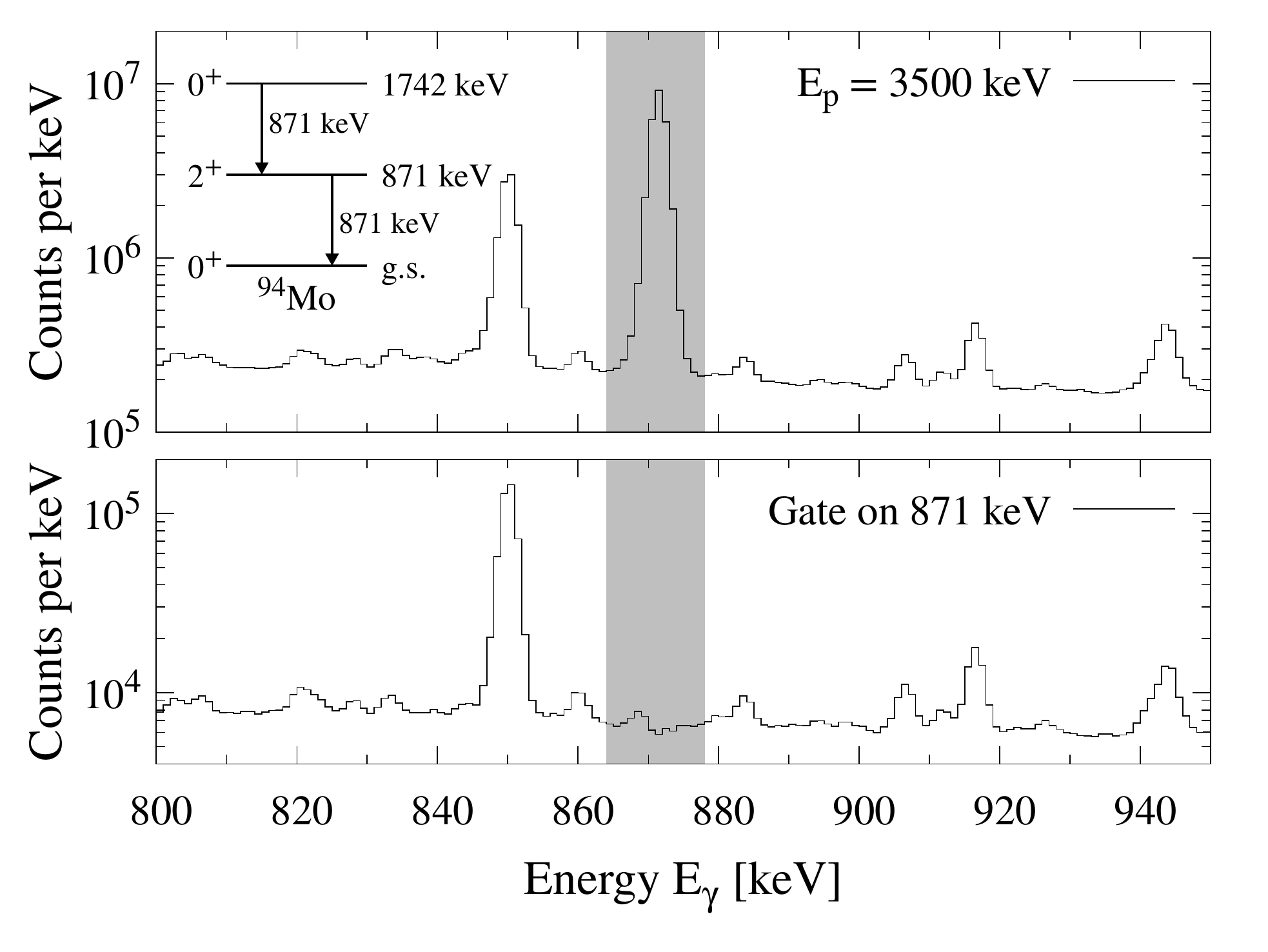}
\caption{The $\gamma\gamma$ coincidence method applied on the $^{93}$Nb(p,$\gamma$)$^{94}$Mo reaction for a proton energy of 3500 keV. Gating on 871 keV in the original spectrum (top panel) reveals that the $0_2^+$ state at 1742 keV is not populated since no coincidences between 871 keV $\gamma$-rays are observed.}  
\label{fig:gate}
\end{figure}

The use of $\gamma\gamma$-coincidences is a well-established and powerful tool to suppress beam-induced background in $\gamma$-ray spectroscopy experiments. Due to the high granularity of HORUS and the event-by-event data format, symmetric $\gamma\gamma$-matrices can be constructed which contain all coincidences between signals of any pair of detectors. This is done using the SOCOv2 (Sorting Code Cologne) \cite{socov2} under consideration of subtracting the time-correlated background.
Using this technique, the background in the $\gamma$-ray spectra can be reduced by many orders of magnitude. Many prompt $\gamma$-rays from the excited compound nucleus have very low intensities and vanish in the background. To determine absolute gamma-ray intensities from $\gamma\gamma$-coincidence spectra the absolute coincidence efficiencies are required. In particular for $\gamma$-ray cascades that contain a prompt $\gamma$-ray from the excited compound state to a low-lying state (in general with $\gamma$-ray energies of up to 10 MeV and more) these cannot be determined experimentally. Since there are no calibration sources which emit $\gamma$-ray cascades including such high $\gamma$-rays and precise information about branching ratios and dead-time of the data acquisition is needed, we can only estimate the coincidence efficiencies from \textsc{Geant4} simulations. 

Applying the $\gamma\gamma$-coincidence method helps to identify unambiguously $\gamma$-ray transitions from the reaction of interest. In the case of the $^{93}$Nb(p,$\gamma$)$^{94}$Mo reaction the $\gamma\gamma$-coincidence technique is very valuable since the $0_2^+$ state at 1742 keV decays via a 871 keV $\gamma$-ray into the $2_1^+$ state which then de-excites via $\gamma$-ray emission with an energy of 871~keV as well. The gate on $E_\gamma=871$ keV (see Fig. \ref{fig:gate}) revealed that the $0_2^+$ state at 1742 keV is not populated in this reaction and the 871~keV peak only contains events from the $2_1^+ \rightarrow g.s.$ transition.

\subsection{Two-Step Cascades}
\label{subsec:TSC}

\begin{figure}[t]
\centering
\includegraphics[width=0.8\columnwidth]{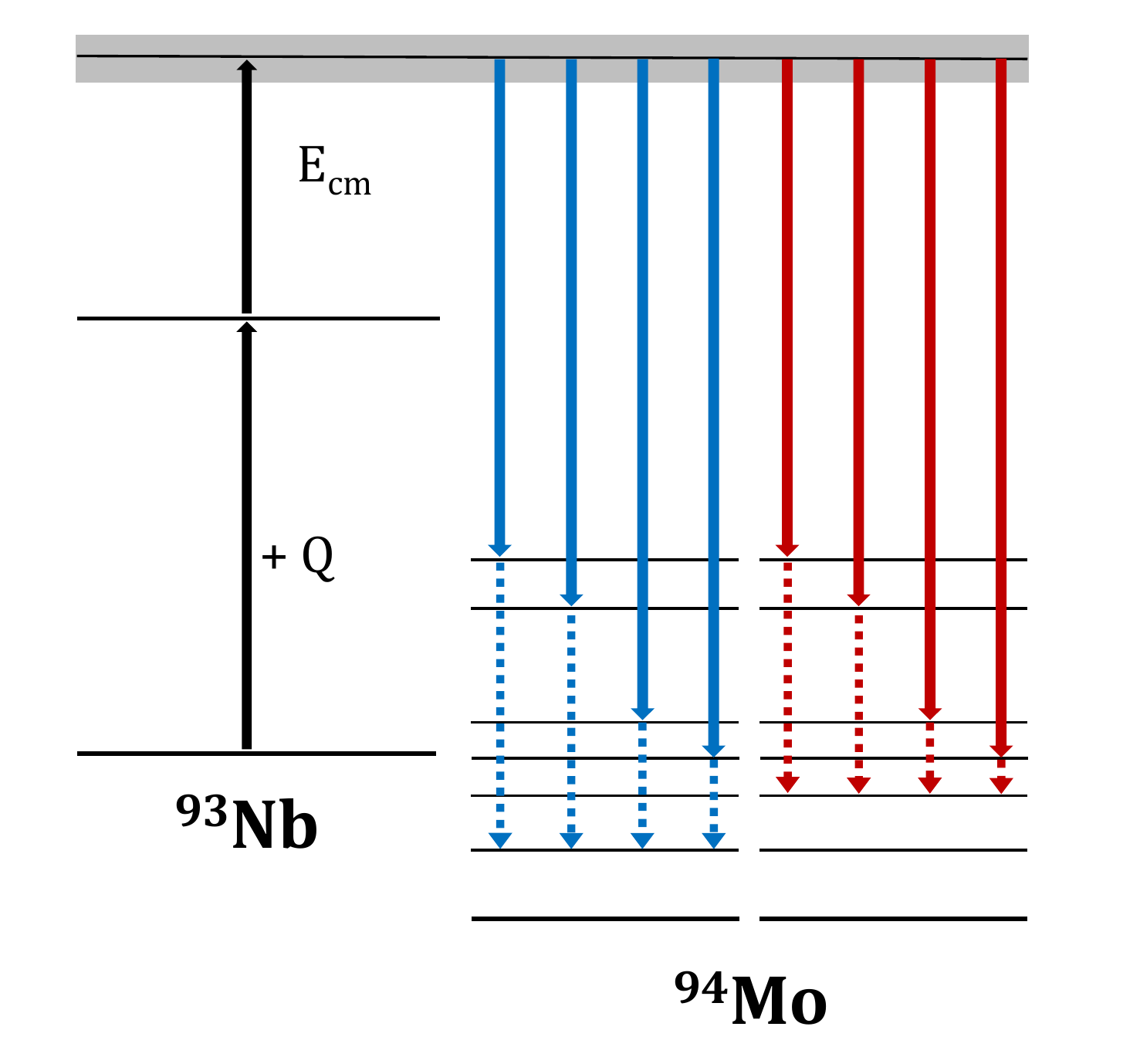}
\caption{Schematic illustration of two step $\gamma$-ray cascades that lead to the population of a specific state. The absolute intensity of high energetic prompt $\gamma$-rays (solid arrows) can be estimated via the intensity of low-energetic \emph{second generation} $\gamma$-rays (dotted arrows). The blue (red) arrows represent the sum peak that contains two $\gamma$-rays which populate the first (second) excited state in the reaction product.}  
\label{fig:tsc}
\end{figure}

Primary $\gamma$-ray transitions are of high interest in most experiments since their intensities are mainly affected by the $\gamma$-ray strength function in the compound nucleus \cite{wiedeking12}. The observation of the corresponding peaks belonging to the prompt $\gamma$-ray transition was already introduced in Sec.\,\ref{subsec:calib}.\ref{subsub:prompt} and very valuable results have been extracted utilizing this method for (p,$\gamma$) reactions on $^{89}$Y, $^{92}$Mo and $^{107}$Ag \cite{Netterdon2015, Mayer2016, Heim2019}.

At those high $\gamma$-ray energies, pair production is a major problem since the resulting single and double-escape peaks in the $\gamma$-ray detectors might overlap with other $\gamma$-ray transitions of interest. In particular, prompt $\gamma$-ray transitions into levels above 2~MeV often cannot be identified unambiguously. This issue can be overcome applying the $\gamma\gamma$ coincidence technique presented in Sec. \ref{subsec:gammagamma}.

Another way of extracting primary $\gamma$-ray transitions is the method of discrete two-step cascade spectra. The TSC method was originally developed to study $\gamma$-ray strength functions and nuclear level densities in (n,$\gamma$) reactions \cite{Becvar1992, Becvar1998, Becvar2007}. TSC matrices can be constructed from the event-by-event listmode format and contain all $\gamma$-ray events that populate a certain level via a two-step $\gamma$-ray cascade from the excited compound state (see Fig. \ref{fig:tsc}). Hence, the sum of these $\gamma$-rays can be described in analogy with Eq. \ref{eq:gamma} with the substitution $E_\gamma=E(\gamma_1)+E(\gamma_2)$. The so-called sum peaks -- the projection of the TSC matrix on the axis of sum energies -- contain events from single $\gamma$-rays which contribute to the population of a certain state. The respective $\gamma$-ray peaks benefit from the high energy resolution of the HPGe detectors since the peaks are not broadened due to energy loss. This method has been succesfully applied to the $^{63,65}$Cu(p,$\gamma$)$^{64,66}$Zn reaction \cite{Scholz2019} and will be applied to future experiments.

\section{Details of the re-designed target chamber}
\label{sec:chamber}

\begin{figure}[t]
\centering
\includegraphics[width=0.9\columnwidth]{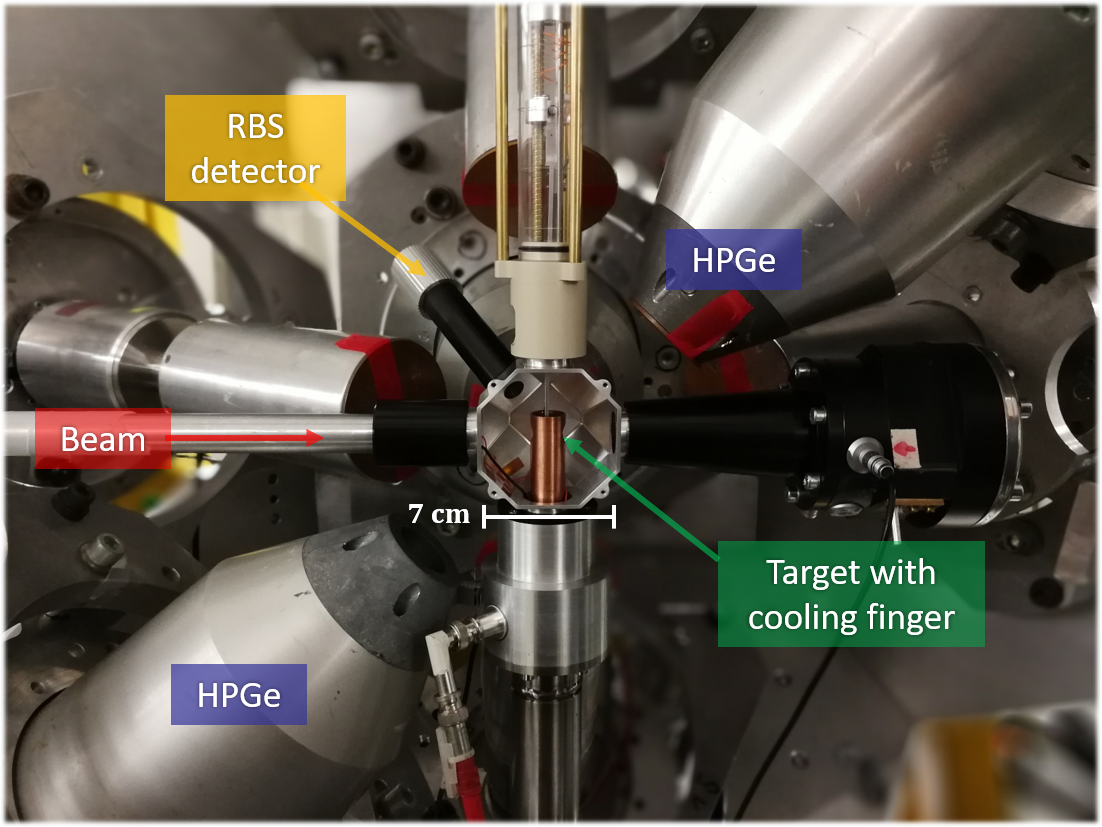}
\caption{The chamber for nuclear astrophysics mounted inside the $\gamma$-ray spectrometer HORUS. The beam (coming from the left side) impinges on the target which is surrounded by a liquid-nitrogen cooled copper finger. The chamber features a much thinner wall and all connectors moved to the stand of the chamber compared to the previous chamber. By this, the overall full-energy peak efficiency increased by a factor of about 2. Furthermore, the chamber is connected to the beam dump and allows to stop the beam at a carbon/tantalum cup a few meters downstream.}  
\label{fig:chamber}
\end{figure}

The target chamber used for nuclear astrophysics experiments at the University of Cologne has been completely revised in 2019 by the in-house mechanical workshop and is shown in Fig. \ref{fig:chamber}. The design of the chamber has changed from a spherical to an asymmetric geometry which is on one side flat and on the other side half of a polyhedron.
The new chamber features a significantly thinner wall of 2\,mm Al and is much more compact in its dimensions compared to the old chamber which is described in Ref. \cite{Netterdon_horus}. The tube that houses the in-beam Rutherford Backscattering Spectrometry (RBS) detector has a length of 9\,cm and a diameter of 2\,cm and its position has been relocated at an azimuthal angle of 45$^{\circ}$. In combination with a reduced volume of the chamber, this allows to mount an additional HPGe detector in HORUS and to move all detectors around 20 \% closer to the target position at distances of around 6 to 13 cm. Additionally, read-outs for the current on the aperture and the chamber as well as the power supply for the suppression voltage were moved to the stand of the chamber. In total, these changes result in a significant increase of the full-energy peak efficiency by a factor of about 2 over the whole energy range (see Fig. \ref{fig:effi}). 

The new chamber can be connected to the beam pipe on both sides. This allows to stop the beam in a carbon or tantalum cup a few meters downstream in experiments in which the beam is not stopped inside the target or its backing and during beam focusing procedures. Compared to the old setup, which required to stop the beam inside or closely behind the chamber, this prevents from activating contaminants close to the target position.
The chamber as well as the beam entrance pipe is coated with 0.1 mm tantalum to reduce beam-induced background further.

\subsection{The in-beam RBS setup}
\label{subsec:target}

\begin{figure}[t]
\centering
\includegraphics[width=1\columnwidth]{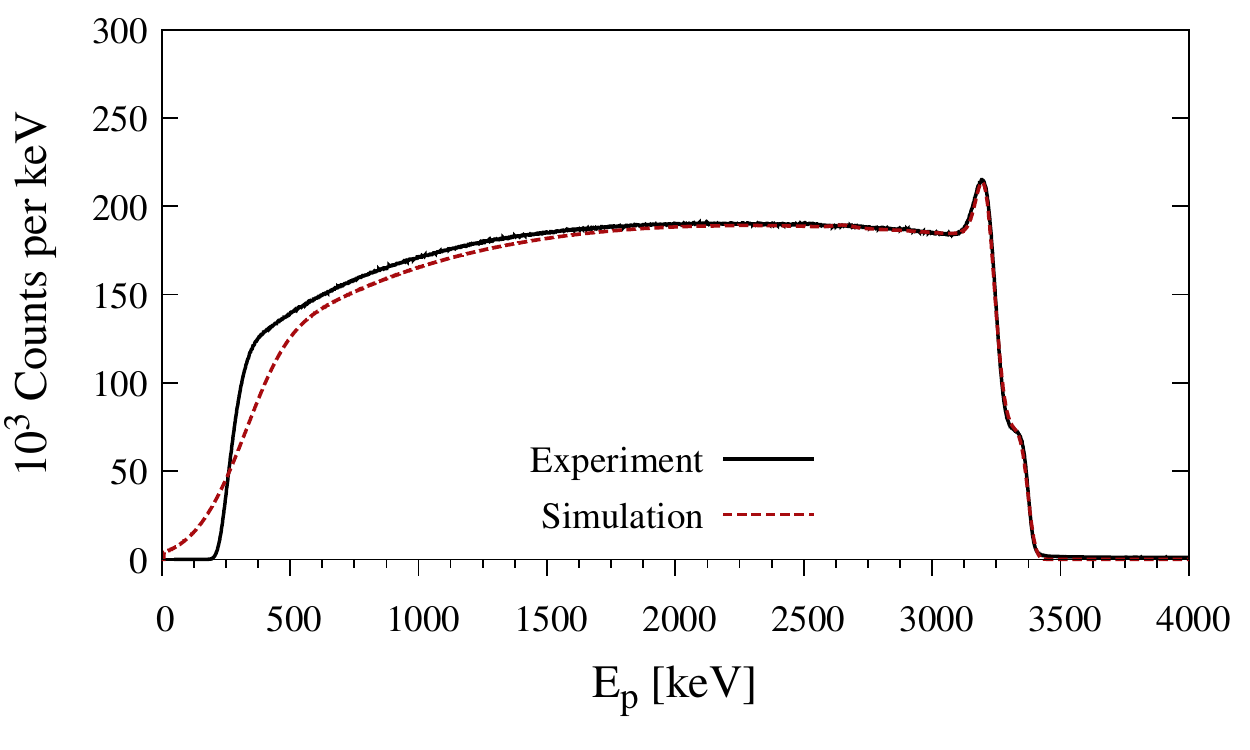}
\caption{RBS spectrum measured with the in-beam silicon detector. Protons with an energy of 2.0 MeV impinged on $^{93}$Nb and a thick gold backing. The edge on the far right side indicates the $^{93}$Nb layer and the next edge the beginning of the $^{197}$Au backing. The falling edge of the peak on top of the plateau determines the thickness of the $^{93}$Nb layer. }  
\label{fig:rbs}
\end{figure}

The built-in silicon detector for Rutherford Backscattering Spectrometry (RBS) measurements is used to monitor the target stability during the experiment and to determine the target thickness. The detector is placed under a backward angle of 135$^\circ$, features an in-beam energy resolution of about 15 keV for protons and covers a solid angle of 0.2 to 20 msr, depending on the diameter of the aperture that is used. A typical RBS spectrum of a $^{93}$Nb target irradiated with protons at an energy of E$_p$~=~3.5 MeV is shown in Fig. \ref{fig:rbs}. The target was manufactured as a self-supporting foil by rolling from metallic niobium. Since niobium is monoisotopic, no other constituents are expected. The targets were very thin -- about 1 mg/cm$^2$ -- in order to minimize the energy loss and thus the width of the prompt $\gamma$-ray peaks (see Fig. \ref{fig:prompt}). The beam is stopped in a gold backing with thickness of around 300~mg/cm$^2$ attached to the backside of the niobium.

The effective target thickness can be extracted from simulations using the SIMNRA code \cite{Mayer02}. The agreement between the measured RBS spectrum and the simulation is excellent (see. Fig. \ref{fig:rbs}).

Typically, the target thickness is additionally determined prior to the experiment using the dedicated RBS setup at the RUBION dynamitron-tandem accelerator at the Ruhr-University Bochum \cite{sauerwein2011, rubion}. Hence, the in-beam RBS setup in the target chamber can be used in two ways.
First, as explained, the in-beam RBS detector can be used to determine the target thickness. The resulting target thickness for $^{93}$Nb (1.3$\pm$0.3 mg/cm$^2$) agrees with the values obtained in Bochum \mbox{(1.1$\pm$0.1 mg/cm$^2$)} within the error bars.
Second, if the target thickness has been determined prior to an experiment, the number of beam particles can be extracted from the in-beam RBS spectra (see Section \ref{subsec:current}).

\subsection{Determination of the number of beam particles}
\label{subsec:current}

To ensure the correct determination of the total number of impinging particles on the target, the beam current is read out at three different positions. The current measurement at the target itself is straightforward, since it is electrically connected to a target ladder. The impinging ion beam on the target leads to scattered beam particles and the release of $\delta$-electrons and therefore the current has to be corrected via measuring the (negative) current on the chamber. Two current integrators determine the accumulated charge individually with an overall uncertainty of about 5 \%. To prevent the $\delta$-electrons from escaping from the chamber into the upstream beampipe, a suppression voltage of \mbox{-400 V} is applied. The current on the Faraday cup behind the chamber is also recorded.

Additionally, the measured in-beam RBS spectra can be used to obtain the total number of beam particles. Given the case that all other relevant input parameters (energy calibration, solid angle of the detector, target thickness) for the RBS simulation explained in Sec. \ref{subsec:target} are sufficiently well determined, the number of beam particles remains the only free parameter. This allows to estimate the beam current with an uncertainty of less than 10 \% and was successfully tested for the $^{93}$Nb(p,$\gamma$)$^{94}$Mo reaction. The simulation shown in Fig. \ref{fig:rbs} yields \mbox{$N_p=(3.6\pm0.3)\,\times10^{10}$} particles and is in excellent agreement with the value obtained from the current read-out which yielded \mbox{$N_p=(3.64\pm0.18)\,\times10^{10}$}.

\subsection{Sensitivity limit}
\label{subsec:sensi}

\begin{table}[t]
\centering

\caption{Relevant parameters that affect the sensitivity limit of in-beam cross-section measurements. During the last years, the experimental performance has significantly improved and has pushed the limit to less than 0.1 $\mu$b. The efficiency at a $\gamma$-ray energy of $E_\gamma$=10\,000~keV is taken into account as well as a typical target thickness of $7 \times 10^{18}$~at/cm$^2$. The experimental values for the irradiation time $t_{beam}$ and beam current $I_{beam}$ are taken from a typical experiment (2014) and using the new setup (2019) respectively. See text for details. }
\vspace{7pt}
\begin{adjustbox}{width=0.8\columnwidth}
\begin{tabular}{c|cccc}
Status &$\epsilon_{eff}$ [\%] & t$_{beam}$ [h] & I$_{proton}$ [nA] & $\sigma_{lim}$ [$\mu$b] \\ \hline \hline
2014	& 0.13 &	70	&	250 &	0.4	\\
2019	& 0.25 &	87	&	685	&	0.05\\
Optimal	& 0.30 &	144	& 1000 &	0.02	\\ \hline
\end{tabular}
\end{adjustbox}
\label{tab:sensitivity}
\end{table} 

The direct measurement of cross sections in the $\mu$b range is very challenging and sets high technical requirements. The sensitivity limit for direct cross-section measurements is affected by the detector efficiency, the maximum available beam current and time. The target thickness is limited by the acceptable energy loss which should not exceed about \mbox{30-40 keV} and is typically in the order of $10^{18}$ at/cm$^2$. In the last years, intensive effort has been put into the improvement of the aforementioned parameters and significantly improved the sensitivity limit for cross-section measurements. The limit has been defined here as the minimum cross section, that yields $\gamma$-ray peaks with at least 1000 counts. Hence, the statistical uncertainty amounts to about 3 \%. The improvement of different parameters during the last five years as well as the the present cross-section limit are given in Table \ref{tab:sensitivity}. Note, that the current lower limit of $\sigma_{min}=0.05$ $\mu$b is very low for an in-beam measurement and has been improved by a factor of 8 during the last five years. The optimal performance in Table \ref{tab:sensitivity} considers a stable beam with an intensity of 1000 nA continuously impinging on the target for 1 week. 

\subsection{Final commissioning}

In order to show that cross section measurements on heavy nuclei using the new setup are reliable, we have measured the $^{89}$Y(p,$\gamma$)$^{90}$Zr reaction as a test case at six different beam energies between $E_p$ = 2 to 3 MeV. We have chosen to use this reaction as commissioning experiment because total cross-section data obtained from three independent experiments exist, see Ref. \cite{Netterdon_horus, tsagari2004, harri2013}. 
We used a natural Y target with an $^{89}$Y enrichment of 99.9 \% which was prepared as a self-supporting foil via rolling. The target thickness was determined from a RBS measurement performed at the Ruhr-University in Bochum (as described in Sec. \ref{subsec:target}) and amounts to 0.85(2) mg/cm$^2$. A 230 mg/cm$^2$ thick gold foil was attached to the backside to ensure that the beam is completely stopped within the target. The irradiation times varied between 30 to 100 minutes (depending on the beam energy) at beam intensities of about 600 nA.

The total (p,$\gamma$) cross section is given by:
\begin{equation} \label{eq:cross}
\sigma_{(p,\gamma)}=\frac{N_{(p,\gamma)}}{N_p \cdot N_t},
\end{equation}
where $N_p$ is the number of projectiles and $N_t$ the number of target nuclei. The total number of reactions $N_{(p,\gamma)}$ is obtained by measuring the angular distribution of all $\gamma$-rays populating the ground state as described in Sec. \ref{subsec:angular}. The combination of Eq. \ref{eq:angular} and \ref{eq:cross} yields the final expression for the total (p,$\gamma$) cross section:
\begin{equation}
\sigma_{(p,\gamma)}=\frac{\Sigma_{i=1}^N\, A_0^i}{N_p \cdot N_t},
\end{equation}
assuming N ground state transitions and $N_{(p,\gamma)}$ as the sum of all $A_0$. For the $^{89}$Y(p,$\gamma$)$^{90}$Zr experiment the intensities of six ground state transitions have been determined. In the compound nucleus $^{90}$Zr, there is a $0_2^+$ state at 1761 keV that decays solely into the ground state via \emph{E0} transition. Since this transition is not observable in the $\gamma$-ray spectra, the population of the $0_2^+$ state at 1761 keV was determined as well. However, for the three lowest beam energies no significant population of the $0_2^+$ state has been observed. Upper values for the contribution of the corresponding transitions were determined and are included in the error of the total cross sections.
In addition, there is an isomeric $5^-$ state at 2319 keV which has a half-life of about 800 ms. In order to detect all $\gamma$-rays from the corresponding decay, the data acquisition has been continued for several seconds after stopping the irradiation.

\begin{figure}[t]
\centering
\includegraphics[width=1\columnwidth]{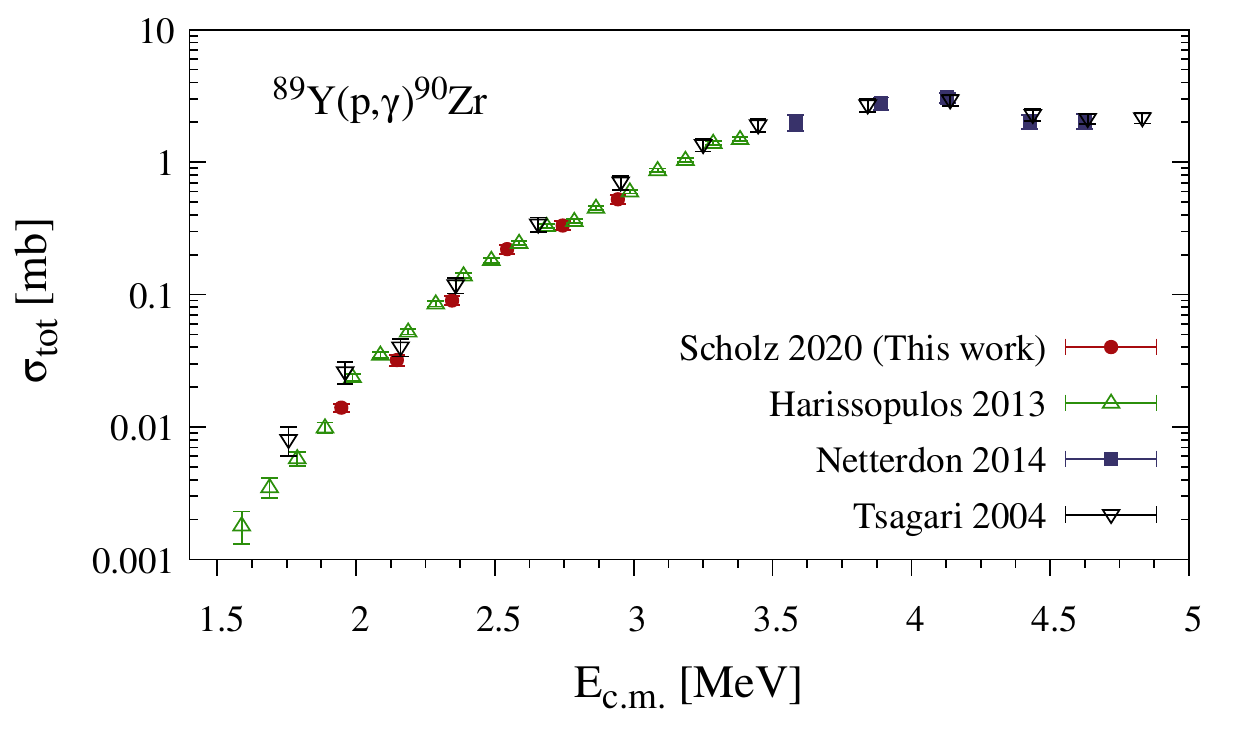}
\caption{Results for the total cross sections of the $^{89}$Y(p,$\gamma$)$^{90}$Zr reaction (red circles) measured with the new setup in Cologne compared to formerly published results \cite{Netterdon_horus, tsagari2004, harri2013}. Since the results are in good agreement with each other, we can conclude that our setup provides reliable data.}  
\label{fig:89y}
\end{figure}

The total cross-section results for the $^{89}$Y(p,$\gamma$)$^{90}$Zr test experiment with our new setup are compared to the existing data in Fig. \ref{fig:89y}. A good agreement is observed, as well as very small uncertainties. In summary, we conclude that the new setup provides robust and reliable data.

\section{Application: Cross-section measurement of the $^{93}$Nb(p,$\gamma$)$^{94}$Mo reaction}
\label{sec:experiments}

Systematic cross-section measurements are the main task of the setup presented in this article. Via the comparison of experimental results to statistical model calculations different nuclear physics input models can either be constrained or excluded. However, the $^{93}$Nb(p,$\gamma$)$^{94}$Mo reaction is of special interest for nuclear astrophysics. The compound nucleus $^{94}$Mo is -- together with a few other $p$ nuclei -- systematically underproduced in various theoretical network calculations \cite{Woosley07, goebel2015}. Therefore, a deeper understanding of the underlying nuclear physics in this nucleus is of paramount importance.

\begin{figure}[t]
\centering
\includegraphics[width=1\columnwidth]{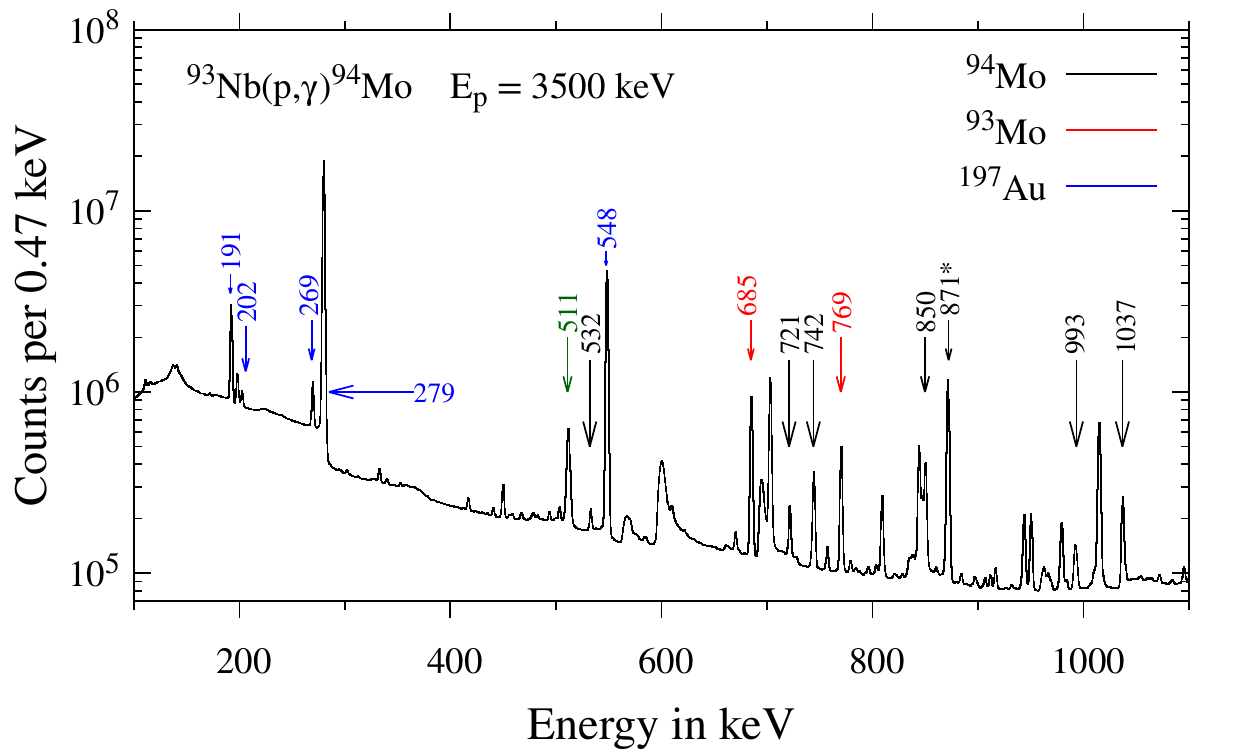}
\caption{Typical $\gamma$-ray spectrum of the $^{93}$Nb(p,$\gamma$)$^{94}$Mo reaction for $E_p$=3.5 MeV. The spectrum was obtained by summing up all detectors mounted under an angle of 90$^{\circ}$ with respect to the beam axis. The $\gamma$-rays that are not produced by the (p,$\gamma$) reaction stem either from elastic scattering on the backing $^{197}$Au or from the $^{93}$Nb(p,n)$^{94}$Mo reaction. The 871 keV transition in $^{94}$Mo is a ground state transition and marked with an asterisk.}  
\label{fig:spectrum}
\end{figure}

The cross section of the $^{93}$Nb(p,$\gamma$)$^{94}$Mo reaction was determined at three proton energies: $E_p$ = 3.0, 3.5 and 4.5 MeV. The Gamow window for this reaction is between 1.71 MeV and 3.39 MeV at a Temperature of 3 GK \cite{Rauscher2010}. Further experimental details were presented in Sec. \ref{subsec:target} and \ref{subsec:current}. Figure \ref{fig:spectrum} shows a typical $\gamma$-ray spectrum for a beam energy of $E_p$=3.5 MeV. The spectrum was obtained by summing up all detectors mounted under an angle of 90$^{\circ}$ with respect to the beam axis. Although considerable beam-induced background, mainly from the $(p,n)$ reaction channel, was observed, all peaks in the spectrum can be clearly identified.

\subsection{Total cross-section results}
\label{subsec:cross}

For $^{93}$Nb(p,$\gamma$)$^{94}$Mo reaction six ground state transitions were observed in total. The largest contribution stems from the $2_1^+ \rightarrow g.s.$ transition at E$_{\gamma}$=871 keV which accounted for about 95 \% of the population of the ground state. Systematic studies have revealed that the relative contributions of various ground-state transitions to the ground-state population remain almost constant with beam energy \cite{Heim2019}.

\begin{table}[b]
\centering
\caption{Total (p,$\gamma$) cross section results for the $^{93}$Nb(p,$\gamma$)$^{94}$Mo reaction. Energies are given as effective center-of-mass energies.}
\vspace{7pt}
\begin{tabular}{c|cccc}
E$_{c.m}$ [keV] & $\sigma_{tot}$ [$\mu$b]  \\ \hline \hline
2988$\pm$10	& 116$\pm$15 	\\
3491$\pm$10	& 227$\pm$31 \\
4495$\pm$10	& 177$\pm$23 	\\ \hline
\end{tabular}
\label{tab:results}
\end{table} 

\begin{figure}[t]
\centering
\includegraphics[width=1\columnwidth]{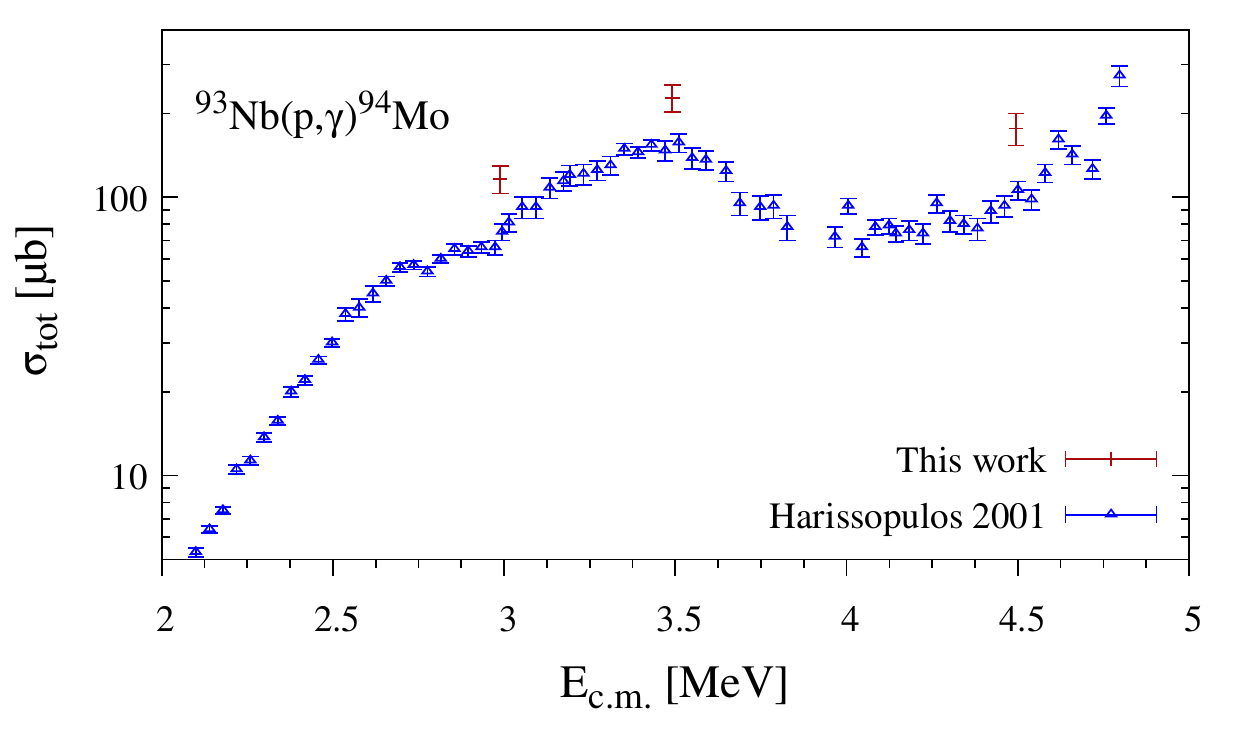}
\caption{Experimental totals cross sections of the $^{93}$Nb(p,$\gamma$)$^{94}$Mo reaction. The results obtained in this work are compared to data from Ref. \cite{harri2002}.}  
\label{fig:totals}
\end{figure}

The cross section results are given in Table \ref{tab:results} and shown in Fig. \ref{fig:totals}. The uncertainties in the cross section values are composed of the uncertainties in the number of projectiles ($\approx$ 5 \%), the target thickness ($\approx$ 10 \%), full-energy peak efficiency ($\approx$ 8 \%) and the statistical error after fitting the Legendre polynomials ($\approx$ 7 \%).
The results obtained in this work are compared to experimental results reported in Ref. \cite{harri2002}. The new results are slightly higher by a factor of about 1.4, which might be explained by the experimental procedure used to derive the results in Ref. \cite{harri2002}.
First, the results from Ref. \cite{harri2002} were obtained by taking only the $2_1^+ \rightarrow g.s.$ transition at E$_{\gamma}$=871 keV into account. In comparison, we have included six ground-state transitions.  
Secondly, in Ref. \cite{harri2002} no suppression voltage is mentioned. Hence, we assume that the measured current could not be corrected for $\delta$-electrons and therefore the cross section is underestimated. 
Third, the angular distributions in Ref. \cite{harri2002} were determined for beam energies between $E_p$=2.0 and 3.0 MeV only and taken as constant over the whole energy range. Additionally, the values above beam energies of $E_p$=3.0 MeV reported in Ref. \cite{harri2002} are scaled to absolute data from a measurement using niobium oxide as a target. For this scaling procedure an additional uncertainty of about 5 \% is reported.

We estimated the possible impact of each of these contributions. Taking correction factor for missing g.s. transitions ($\approx$ 6 \%), unobserved $\delta$-electrons ($\approx$ 10 \%), wrong angular correlations ($\approx$ 18 \%) and uncertainties from the scaling procedure ($\approx$ 5 \%) into account might explain the discrepancy.

\section{Summary}
\label{sec:conclusions}

In this article the experimental setup for experiments to study cross sections relevant for nuclear astrophysics at the University of Cologne was presented. The HORUS $\gamma$-ray spectrometer combined with the revised target chamber is an excellent tool to measure even very small cross sections down to the nb range at astrophysically relevant energies. Additionally, it enables the measurement of $\gamma\gamma$-coincidences and discrete two-step cascade spectra which allow to perform an even more sophisticated analysis and study important nuclear physics concepts as the $\gamma$-ray strength function or nuclear level densities.

The astrophysical relevant $^{93}$Nb(p,$\gamma$)$^{94}$Mo reaction was successfully measured and total (p,$\gamma$) cross sections for proton energies between $E_p=3.0 - 4.5$ MeV have been extracted. The results are compared to previously measured data from Ref. \cite{harri2002}. 

In summary, the nuclear astrophysics setup in Cologne allows to measure cross sections of reactions with stable and unstable compound nuclei and will further help to extend the available experimental database and study the underlying nuclear physics parameters that enter theoretical nucleosynthesis network calculations.

\section{Outlook}
\label{sec:outlook}

In addition to the 10 MV FN-Tandem accelerator, the University of Cologne operates a 6 MV Tandetron accelerator which is primarily used for Accelerator Mass Spectrometry (AMS) \cite{CologneAMS11}. The accelerator can provide stable proton beam intensities of several $\mu$A with very low ripple. At present, construction of a HPGe array which is solely dedicated for nuclear astrophysics experiments is being planned for this machine. Seven HPGe detectors with relative efficiencies of 80\,\% and two clover detectors with relative efficiencies of 120\,\% will be mounted. The Tandetron accelerator can be operated independently from the 10 MV FN-Tandem accelerator. Commissioning of this setup is expected to begin in the near future.

\section*{Acknowledgments}
We gratefully thank K. O. Zell and A. Blazhev for the target preparation, and H.W. Becker and V. Foteinou of the Ruhr-Universit\"at Bochum for the assistance during the RBS measurements.
This project has been supported by the Deutsche Forschungsgemeinschaft under the contracts ZI 510/8-1.

\begin{thebibliography}{00}



\bibitem{bbfh} E. Burbidge, G. Burbidge, W. Fowler, F. Hoyle, Rev. Mod Phys. \textbf{29} (1957) 547.

\bibitem{Arnould2003} M. Arnould, S. Goriely, Phys. Rep. \textbf{384} (2003) 1.

\bibitem{Rauscher2013} T. Rauscher, N. Dauphas, I. Dillmann, C. Fr\"ohlich, Zs. F\"ul\"op, and Gy. Gy\"urky, Rep. Prog. Phys. 76 (2013) 066201.

\bibitem{Gyurky2019} Gy. Gy\"urky, Zs. F\"ul\"op, F. K\"appeler, G.G. Kiss, and A. Wallner, Eur. Phys. J. A \textbf{55} (2019) 41.

\bibitem{Netterdon2014} L. Netterdon, A. Endres, G.G. Kiss, J. Mayer, T. Rauscher, P. Scholz, K. Sonnabend, Z. T\"or\"ok, A. Zilges, Phys. Rev. C \textbf{90} (2014) 035806.

\bibitem{Gyurky2014} G. Gy\"urky, Z. F\"ul\"op, Z. Hal\'{a}sz, G.G. Kiss, T. Sz\"ucs, Phys. Rev. C \textbf{90} (2014) 052801.

\bibitem{Gyuky2014_2} G. Gy\"urky, M. Vakulenko, Zs. F\"ul\"op, Z. Hal\'{a}sz, G. Kiss, E. Somorjai, T. Sz\"ucs, Nucl. Phys. A \textbf{922} (2014) 112.

\bibitem{Kiss2014} G. Kiss, T. Sz\"ucs, T. Rauscher, Z. T\"or\"ok, Z. F\"ul\"op, Gy. Gy\"urky, Z. Hal\'{a}sz, E. Somorjai, Phys. Lett. B \textbf{735} (2014) 40.

\bibitem{Scholz2014} P. Scholz, A. Endres, A. Hennig, L. Netterdon, H.W. Becker, J. Endres, J. Mayer, U. Giesen, D. Rogalla, F. Schlüter \textit{et al.}, Phys. Rev. C \textbf{90} (2014) 065807.

\bibitem{Guray2015} R.T. G\"uray, N. \"Ozkan, C. Yal\c{c}in, T. Rauscher, G. Gy\"urky, J. Farkas, Z. F\"ul\"op, Z. Hal\'{a}sz, E. Somorjai, Phys. Rev. C \textbf{91} (2015) 055809.

\bibitem{Kiss2015} G.G. Kiss, T. Sz\"ucs, T. Rauscher, Z. Török, L. Csedreki, Z. F\"ul\"op, G. Gy\"urky, Z. Hal\'{a}sz, J. Phys. G \textbf{42} (2015) 055103.

\bibitem{Yalcin2015} C. Yal\c{c}in, G. Gy\"urky, T. Rauscher, G.G. Kiss, N. \"Ozkan, R.T. G\"uray, Z. Hal\'{a}sz, T. Sz\"ucs, Z. Fülöp, J. Farkas et al., Phys. Rev. C \textbf{91} (2015) 034610.

\bibitem{Halasz2016} Z. Hal\'{a}sz, E. Somorjai, G. Gy\"urky, Z. Elekes, Z. F\"ul\"op, T. Sz\"ucs, G.G. Kiss, N.T. Szegedi, T. Rauscher, J. G\"orres et al., Phys. Rev. C \textbf{94} (2016) 045801.

\bibitem{Kinoshita2016} N. Kinoshita, K. Hayashi, S. Ueno, Y. Yatsu, A. Yokoyama, N. Takahashi, Phys. Rev. C \textbf{93} (2016) 025801.

\bibitem{Scholz2016} P. Scholz, F. Heim, J. Mayer, C. M\"unker, L. Netterdon, F. Wombacher, A. Zilges, Phys. Lett. B \textbf{761} (2016) 247.

\bibitem{Ozkan2017} N. \"Ozkan, R.T. G\"uray, C. Yal\c{c}in, W.P. Tan, A. Aprahamian, M. Beard, R.J. deBoer, S. Almaraz-Calderon, S. Falahat, J. G\"orres et al., Phys. Rev. C \textbf{96} (2017) 045805.

\bibitem{Spyrou07} A. Spyrou, H.-W. Becker, A. Lagoyannis, S. Harissopulos, and C. Rolfs, Phys. Rev. C \textbf{76} (2007) 015802.

\bibitem{Spyrou08} A. Spyrou, A. Lagoyannis, P. Demetriou, S. Harissopulos, and H.-W. Becker, Phys. Rev. C \textbf{77} (2008) 065801.
\bibitem{Simon13} A. Simon \emph{et al.}, Nucl. Instr. Meth. A \textbf{703} (2013) 16.
\bibitem{Simon19} A. Simon \emph{et al.} J. Phys.: Conf. Ser. \textbf{1308} (2019) 012020.
\bibitem{reingold19} C.S. Reingold \emph{et al.}, Eur. Phys. J. A \textbf{55}, (2019) 77.
\bibitem{Netterdon_horus} L. Netterdon, V. Derya, J. Endres, C. Fransen, A. Hennig, J. Mayer, C. M\"uller-Gatermann, A. Sauerwein, P. Scholz, M. Spieker, A. Zilges, Nucl. Instr. and Meth. A \textbf{754} (2014) 94.
\bibitem{Netterdon_Sn} L. Netterdon, J. Mayer, P. Scholz, A. Zilges, Phys. Rev. C \textbf{91} (2015) 035801.
\bibitem{Hennig15} A. Hennig, V. Derya, M.N. Mineva, P. Petkov, S.G. Pickstone, M. Spieker, A. Zilges, Nucl. Instr. Meth. A \textbf{794} (2015) 171.
\bibitem{pickstone2017} S. G. Pickstone, M. Weinert, M. F\"arber, F. Heim, E. Hoemann, J. Mayer, M. M\"uscher, S. Prill, P. Scholz, M. Spieker, V. Vielmetter, J. Wilhelmy, and A. Zilges, Nucl. Instr. Meth. A \textbf{874} (2017) 104.

\bibitem{Spieker18} M. Spieker, P. Petkov, E. Litvinova, C. M\"uller-Gatermann, S.G. Pickstone, S. Prill, P. Scholz, A. Zilges, Phys. Rev. C \textbf{97} (2018) 054319.


\bibitem{Hubbard-Nelson99} B. Hubbard-Nelson, M. Momayezi, and W. Warburton, Nucl. Instr. Meth. A \textbf{422} (1999) 411.

\bibitem{Skulski00} W. Skulski, M. Momayezi, B. Hubbard-Nelson, P. Grudberg, J. Harris, and W. Warburton, Acta Phys. Pol. B \textbf{31} (2000) 47.

\bibitem{Hennig14} A. Hennig, C. Fransen, W. Hennig, G. Pascovici, N. Warr, M. Weinert, A. Zilges, Nucl. Instr. Meth. A \textbf{758} (2014) 69.
\bibitem{Rauscher2010} T. Rauscher, Phys. Rev. C \textbf{81} (2010) 045807.

%
\bibitem{Brenneisen1995} J. Brenneisen, D. Grathwohl, M. Lickert, R. Ott, H. R\"opke, J. Schm\"alzlin, P. Siedle, and B.~H. Wildenthal, Z. Phys. A \textbf{352} (1995) 149.
%
\bibitem{Srim} J.~F. Ziegler and J.~P. Biersack, Code \textsc{Srim}, version 2012.03, full description given by J.~F. Ziegler, J.~P. Biersack, and U. Littmark, \textit{The Stopping and Range of Ions in Solids} (Pergamon Press, New York, 1985).

\bibitem{g4Horus} J. Mayer, \emph{G4Horus - Geant4 Simulation for Horus}, https://gitlab.ikp.uni-koeln.de/jmayer/g4horus.git, (2019).

\bibitem{66Ga} E. Browne, J.K. Tuli, Nuclear Data Sheets \textbf{111} (2010) 1093.
\bibitem{socov2} Nima Saed-Samii, \emph{SOCOv2 - Sorting Code Cologne}, https://gitlab.ikp.uni-koeln.de/nima/soco-v2.git (2019).

\bibitem{Geant4} N. Agostinelli \textsl{et al.}, Nucl. Instr. Meth. A \textbf{506} (2003) 250.

\bibitem{wiedeking12} M. Wiedeking \emph{et al.}, Phys. Rev. Lett. \textbf{108} (2012) 162503.

\bibitem{Netterdon2015} L. Netterdon, A. Endres, S. Goriely, J. Mayer, P. Scholz, M. Spieker, and A. Zilges, Phys. Lett. B \textbf{744} (2015) 358.

\bibitem{Mayer2016} J. Mayer, S. Goriely, L. Netterdon, S. P\'{e}ru, P. Scholz, R. Schwengner, and A. Zilges, Phys. Rev. C \textbf{93} (2016) 045809.

\bibitem{Heim2019} F. Heim, P. Scholz, M. K\"orschgen, J. Mayer, M. M\"uller, and A. Zilges, Phys. Rev C \textbf{101} (2020) 035805. 

\bibitem{Becvar1992} F. Be\v{c}v\'{a}\v{r}, P. Cejnar, R. E. Chrien, and J. Kopeck\'{y}, Phys. Rev. C \textbf{46} (1992) 1276.

\bibitem{Becvar1998} F. Be\v{c}v\'{a}\v{r}, Nucl. Instr. Meth. A \textbf{417} (1998) 434.

\bibitem{Becvar2007} F. Be\v{c}v\'{a}\v{r} \emph{et al.}, Nucl. Instr. Meth. B \textbf{261} (2007) 930.

\bibitem{Scholz2019} P. Scholz, M. Guttormsen, F. Heim, A. C. Larsen, J. Mayer, D. Savran, M. Spieker, G. M. Tveten, A. V. Voinov, J. Wilhelmy, F.  Zeiser, and A. Zilges, Phys. Rev. C \textbf{101} (2020) 045806.

\bibitem{Mayer02} M. Mayer, Nucl. Instr. Meth. B \textbf{194} (2002) 177.

\bibitem{sauerwein2011} A. Sauerwein \emph{et al.}, Phys Rev. C \textbf{84} (2011) 045808.
\bibitem{rubion} \mbox{RBS - Rutherford Backscattering Spectrometry RUBION}, https://www.rubion.rub.de/en/methods/rbs-rutherford-backscattering-spectrometry/

\bibitem{Woosley07} S.~E. Woosley and A. Heger, Phys. Rep. \textbf{442} (2007) 269.

\bibitem{goebel2015} K. G\"obel, J. Glorius, A. Koloczek, M. Pignatari, R. Reifarth, R. Schach und K. Sonnabend, EPJ Web of Conferences \textbf{93} (2015) 03006.
\bibitem{harri2002} S. Harissopulos \emph{et al.}, Phys. Rev. C \textbf{64} (2001) 055804.

\bibitem{martini2014} M. Martini, S. Hilaire, S. Goriely, A.J. Koning, and S. P\'eru, Nucl. Data Sheets \textbf{118} (2014) 273.
\bibitem{goriely2015} S. Goriely, Eur. Phys. J. A \textbf{51} (2015) 172.

\bibitem{CologneAMS11} M.G. Klein, A. Dewald, A. Gottdang, S. Heinze, D.J.W. Mous, Nucl. Instr. Meth. B \textbf{269} (2011) 3167.

\bibitem{tsagari2004} P. Tsagari, M. Kokkoris, E. Skreti, A. G. Karydas, S. Harissopulos, T. Paradellis, and P. Demetriou, Phys. Rev C \textbf{70} (2004) 015802.

\bibitem{harri2013} S. Harissopulos, A. Spyrou, A. Lagoyannis, M. Axiotis, P. Demetriou, J. W. Hammer, R. Kunz, and H.-W. Becker, Phys. Rev. C \textbf{87} (2013) 025806.

%
%


\end{thebibliography}



\end{document}